\documentclass[10pt,conference]{IEEEtran}
\IEEEoverridecommandlockouts
\usepackage{cite}
\usepackage{amsmath,amssymb,amsfonts}
\usepackage{algorithmic}
\usepackage{graphicx}
\usepackage{textcomp}
\usepackage{xcolor}
\usepackage{cleveref}
\usepackage[utf8]{inputenc}
\usepackage{url}
\usepackage{soul}
\usepackage{subfigure}
\usepackage{ulem}
\usepackage{showexpl}
\usepackage{inconsolata}

\usepackage[compact]{titlesec}  

\titlespacing{\section}{4pt}{2pt}{2pt}

\AtBeginDocument{
 \setlength\abovedisplayskip{2pt}
  \setlength\belowdisplayskip{0pt}
  }

\title{Cyber Framework for Steering and Measurements Collection Over Instrument-Computing Ecosystems
\thanks{This research is sponsored in part by the INTERSECT Initiative as part of the Laboratory Directed Research and Development Program and in part by RAMSES project of Advanced Scientific Computing Research program, U.S. Department of Energy, and in part by the Office of Basic Energy Sciences, Division of Materials Sciences and Engineering, U.S. Department of Energy, and is performed at Oak Ridge National Laboratory managed by UT-Battelle, LLC for U.S. Department of Energy under Contract No. DE-AC05-00OR22725.
The United States Government retains and the publisher, by accepting the article for publication, acknowledges that the United States Government retains a nonexclusive, paid-up, irrevocable, world-wide license to publish or reproduce the published form of this manuscript, or allow others to do so, for United States Government purposes. The Department of Energy will provide public access to these results of federally sponsored research in accordance with the DOE Public Access Plan (http://energy.gov/downloads/doe-public-access-plan).
}
}

\author{
\IEEEauthorblockN{
Anees Al-Najjar, Nageswara S. V. Rao, Ramanan Sankaran, Helia Zandi,\\ Debangshu Mukherjee, Maxim Ziatdinov, Craig Bridges}
\IEEEauthorblockA{
\textit{Oak Ridge National Laboratory}\\ Oak Ridge, TN, USA\\
\{alnajjaram,raons,sankaranr,zandih,mukherjeed,ziatdinovma,bridgesca\}@ornl.gov}\\
}

\begin{document}

\maketitle
\begin{abstract}
We propose a framework to develop cyber solutions to  support the remote steering of science instruments and measurements collection over instrument-computing ecosystems. It is based on provisioning separate data and control connections at the network level, and developing software modules consisting of Python wrappers for instrument commands and Pyro server-client codes that make them available across the ecosystem network.
We demonstrate automated measurement transfers and remote steering operations in a microscopy use case for materials research over an ecosystem of Nion microscopes and computing platforms connected over site networks. The proposed framework is  currently under further refinement and being adopted to science workflows with automated remote experiments steering 
for autonomous chemistry laboratories and smart energy grid simulations.
\end{abstract}


\section{introduction}
\label{sec:introduction}

Distributed workflows orchestrated by artificial intelligence (AI) over science instrument-computing ecosystems (ICE) hold an enormous promise for accelerating the productivity and discovery in 
science scenarios. 
Their success depends on the ability to orchestrate automated experiments at remote physical instruments by steering them and collecting the generated measurements. 
We consider physical science 
{ instruments and computing platforms existing at geographically dispersed locations } that form  an {ecosystem} with a goal to seamlessly support these workflows \cite{9307775}.
The underlying tasks may involve configuring instruments, collecting and transferring measurements, and analyzing them at remote computing systems to extract parameters for the next step in a series of experiments.
{Nowadays, a number of these tasks are performed manually} as part of a workflow that may span days to weeks where experiments are repeated using different parameters produced by simulation and analysis codes.
These human-driven operations {confine the scalability and performance of scientific workflows, leading to inadequate utilization of high-cost instruments.} 

\begin{figure}
\centering
\includegraphics[width=0.5\textwidth]{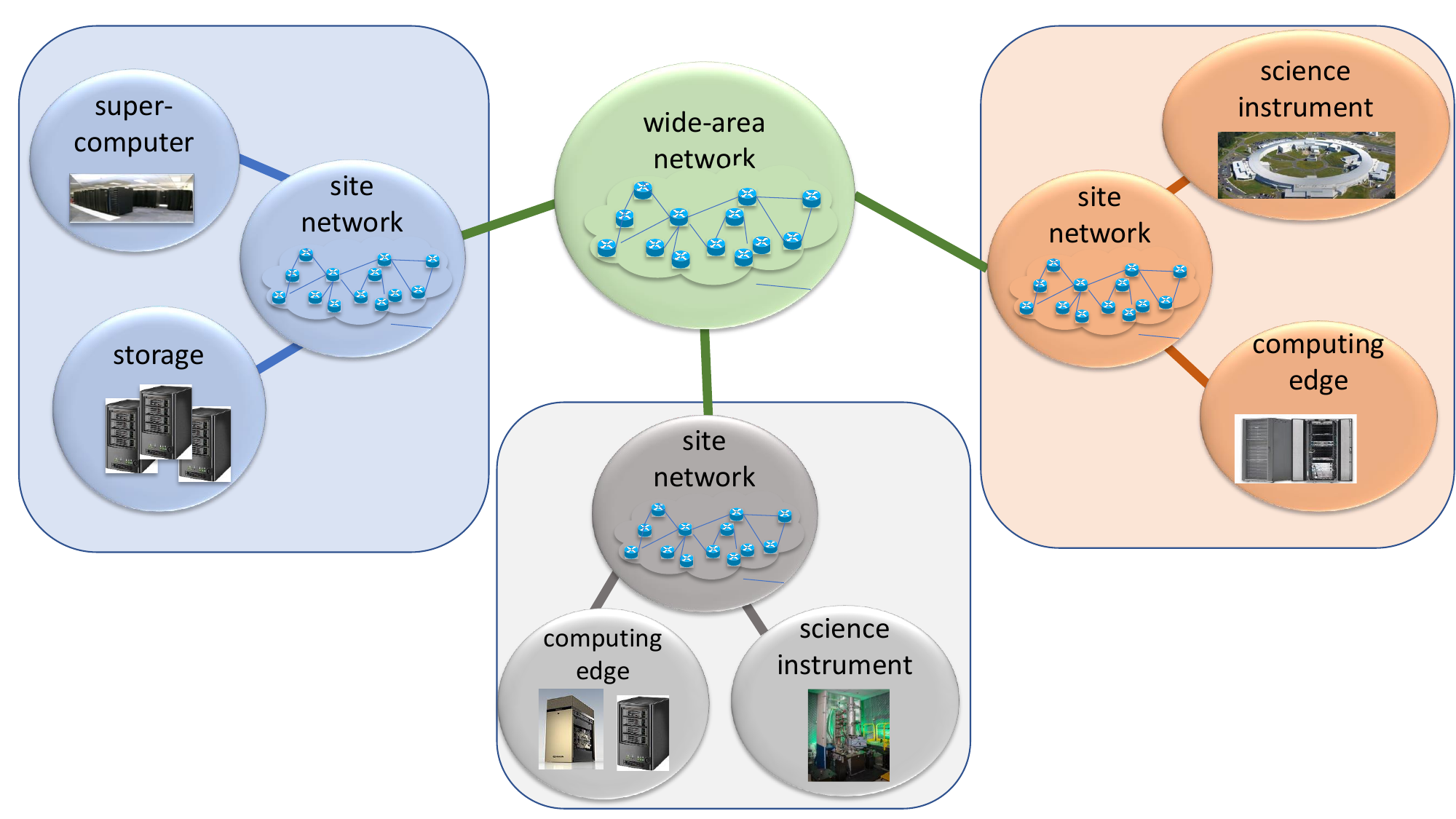}
\caption{\small A complex ecosystem of science instruments, computing and storage sites connected over a wide-area network
.} 
\label{fig:science_complex_ecosystem}
\vspace*{-0.2in}
\end{figure}

{ The science workflows may require computing and storage services provisioned by complex science ecosystems} 
 that utilize diverse instruments, for example, light sources \cite{bicer2017real}, chemistry instruments
, microscopes \cite{4dstem_review}, and others.
{ Generally, these instruments are locally controlled through Windows-based platforms, while the remote computing systems are GPU Linux-based situated at different networked domains separated by firewalls. }
{So far,  a general cyber framework does not exist to support steering and measurement collections needed for these ICE workflows with diverse instruments, except solutions deployed at neutron and light source facilities, like the Experimental Physics and Industrial Control System (EPICS) \cite{epics}.
To fully realize the potential of diverse ICE,  the design, implementation and testing of software modules are mandated to collect and transfer the acquired data, as well as perform cross-facility instrumentation steering, as illustrated in Fig.~\ref{fig:science_complex_ecosystem}.}

In addition, the underlying network connections need to be provisioned to support custom, protected flows needed for concurrent steering and data transfer operations.

\begin{figure}
\centering
\includegraphics[width=0.5\textwidth]{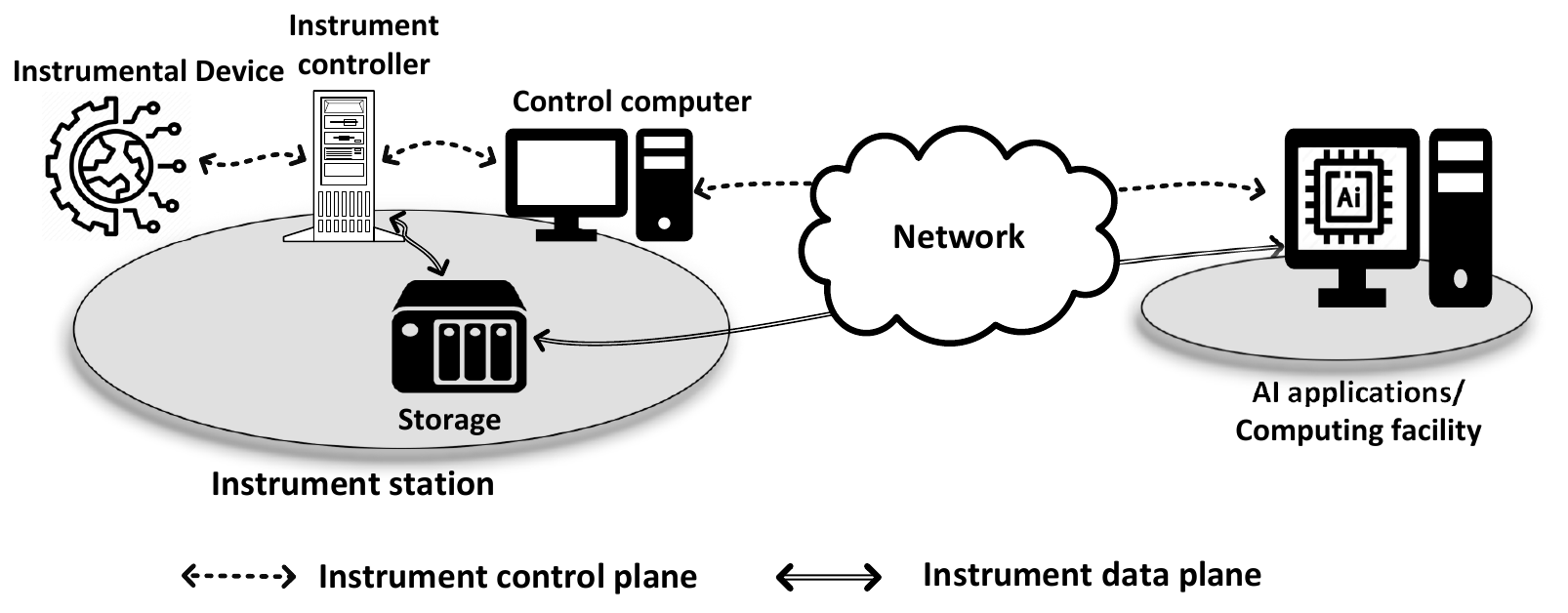}
\caption{\small Cyber framework for implementing remote steering and measurements collection capabilities over
an ecosystem.} 
\label{fig:Cyber_framework}
\vspace*{-0.1in}
\end{figure}

We propose a cyber framework for implementing remote steering and measurements collection capabilities over ICE, as illustrated in Fig.~\ref{fig:Cyber_framework}. 
Separate data and control connections are set up at the network level by configuring the physical ecossystem connections.
Software modules are developed by wrapping instrument APIs and control commands in Python, and exposing them using Pyro client-server codes for network access. 
This framework is general and applicable to instruments that provide APIs (both Windows and Linux) and network interfaces.

We describe the design and implementation of this framework, and workflow execution using Jupyter notebook in Section~\ref{sec:design_and_implementation}.
{ We illustrate automatic data transfer and cross-facility control} over ORNL site networks for microscopy workflows in~\ref{sec:experimental_setup}.

\section{Cyber Framework: Design and Implementation}
\label{sec:design_and_implementation}

We propose a cyber framework for implementing remote steering and measurements collection capabilities over ICE, illustrated in Fig.~\ref{fig:Cyber_framework}. 
Scalable and reliable science workflows are supported utilizing a novel cyber framework of network connections and software modules, explained as follows. 
\begin{itemize}
\item [(a)] {\textit{Network connections:}} Separate data and control connections are set up at the network level by configuring physical connections between ecosystem components at different facilities. These cross-facility connections are established by aligning the firewalls and access policies of networks, facilities, and hardware and system applications at edge systems, to grant network traffic access.
\item [(b)] {\textit{Cross-platform software modules:}} 
%
Software modules are developed to support cyber-physical and analytical operations, including instruments control (typically run on Windows platforms), measurements analytics (typically run on mutli-GPU Linux systems), and cross-facility measurements transfer. The instruments control modules utilize Python to wrap instruments APIs and control commands to be exposed and called via Pyro client-server modules for remote instruments steering possibly by AI codes. The data transfer is enabled across the ecosystem between Windows-based storage, such as network-attached storage (NAS), located near the instrument and remote Linux systems, as shown in Fig.~\ref{fig:Cyber_framework}. 
We utilize cross-platform file mounting methods specific to facility setup, for example,  Secure Shell File System (SSHFS) between Windows and Linux platforms.
\end{itemize}

\begin{figure}[thb]
\centering
\includegraphics[width=0.52\textwidth]{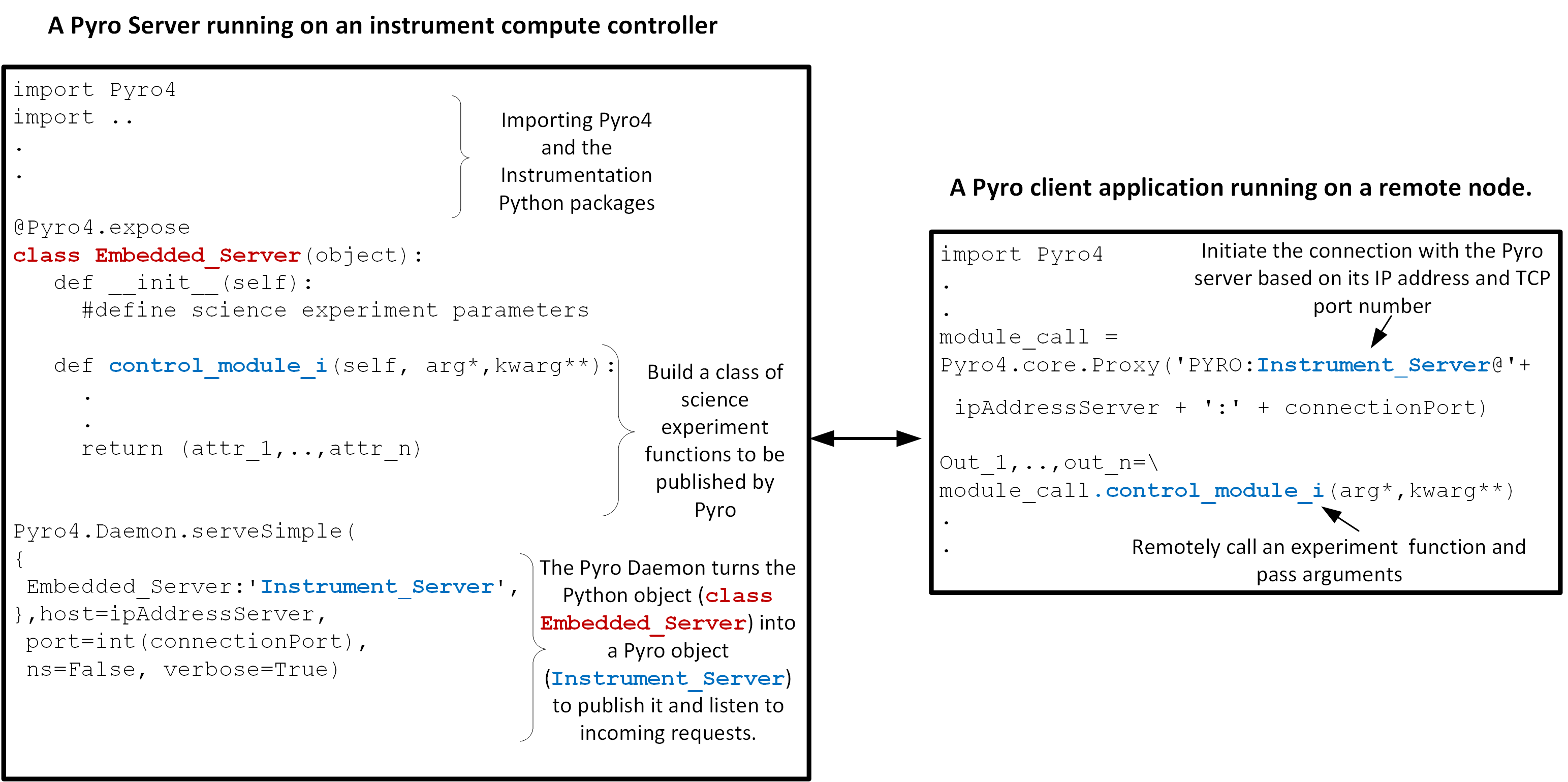}
\caption{Pyro client-server communication applied between the instrument control node and a remote compute system.} 
\label{fig:Pyro_server_client_setup}
\end{figure}

\subsection{Control Channel}
\label{subsec:Control Connections}

 Steering science experiments and instruments (via their control computers) across an ecosystem is accomplished through remote control commands sent via a (network) control channel. The incorporated control computers respond back, as a result of the execution of the control commands, with measurements and possibly metadata and system messages. We developed Pyro client-server codes to support remote steering of such experiments across the ecosystem. Pyro provides a Python API for network access~\cite{pyro}.
We developed and installed Pyro server modules on the control computer, as part of instrument control software
, while the client modules are installed on remote computing systems.

{ The client-server communication over a multi-site ecosystem is explained in Fig.~\ref{fig:Pyro_server_client_setup}. The Pyro modules at the client access the controls and programmable interfaces on the server. The client modules are run as part of automated workflow from a console or a web-based interactive platform, such as a Jupyter Notebook. The communication passes the control node IP address with the control commands and parameters required to steer the instruments. The proposed framework supports the concurrent execution of Pyro client modules on multiple remote systems, which communicate with the Pyro server to execute the exposed control commands on the control platforms.}

\subsection{Data Channel}
\label{subsec:Data Connections}

 The data channel allows the acquired measurements at a science facility, as a result of the execution of the instrumentation control commands, to be shared and available at the remote servers across the ecosystem.
Usually, instrument controller APIs are configured to store the measurements on Windows-based storage as files, as in the case of Swift software (Nion microscopes controller API), that stores the Scanning transmission electron microscopy (STEM) measurements on NAS system.

We implement the data channel by remotely cross-mounting the data files using a certain file sharing technique  (Common Internet File System (CIFS) or SSHFS) for providing access across different operating systems
. The technique depends on:  (i) setup of instrument storage and remote computing systems, and (ii) network domain configurations and traffic access policies on the incorporated systems.
 The file-sharing access privileges on the computing nodes are configured for granting access to authorized users to the measurement files across the ecosystem. The access configuration is performed once for permanent file sharing across the ecosystem.

\subsection{Networks, Access and Firewalls}
\label{subsec:Networks_Access_and_Firewalls}

The ecosystem components of computers and instruments dispersed over multiple facilities are integrated into various domains isolated with network and system host firewalls. To enable the data channel, firewall rules are inserted to procure file mounting between the storage and computing servers through the data channel. Firewall rules are also injected to open communication ports for Pyro servers and clients over the control channel.


\section{experimental setup and Demonstration}
\label{sec:experimental_setup}

The ecosystem capabilities described in the previous section are implemented on Oak Ridge National Laboratory (ORNL) ecosystem comprising of scientific instruments and remote computing systems. The microscopy workflow utilizes Nion microscopes located at the Center for Nanophase Materials Sciences (CNMS). The computing facility utilizes GPU-based Linux systems available at K200 computing facility located in a different building. We utilize the proposed cyber framework design, discussed in Section~\ref{sec:design_and_implementation}, to integrate the science instruments with remote computing systems across ORNL network to perform remote instrument steering and measurement transfers.

\begin{figure}[t]
\centering

\includegraphics[width=0.45\textwidth]{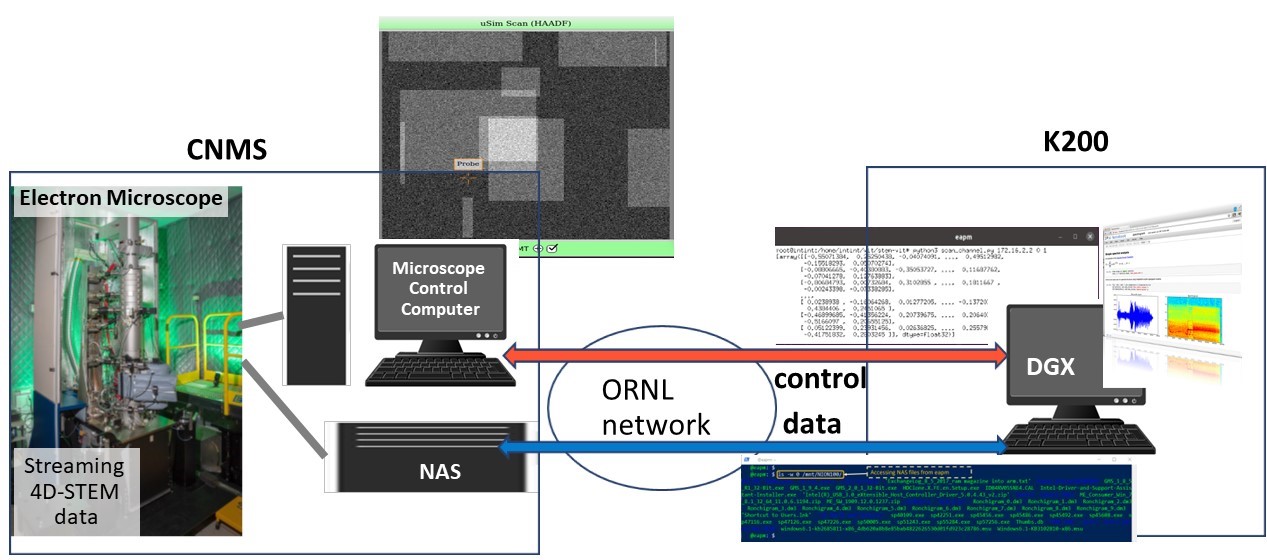}
\caption{Cross-facility integration with data and control channels between U200 at CNMS and DGX at K200 over ORNL  infrastructure.} 
\label{fig:Remote steering Microscopy experiments}
\end{figure}

This workflow uses U200 Nion microscope with its NAS at CNMS, and DGX workstation at the K200 computing facility as part of ORNL physical infrastructure shown in  Fig.~\ref{fig:Remote steering Microscopy experiments}.

\begin{figure}[t]
	\centering
\includegraphics[width=0.9\linewidth]{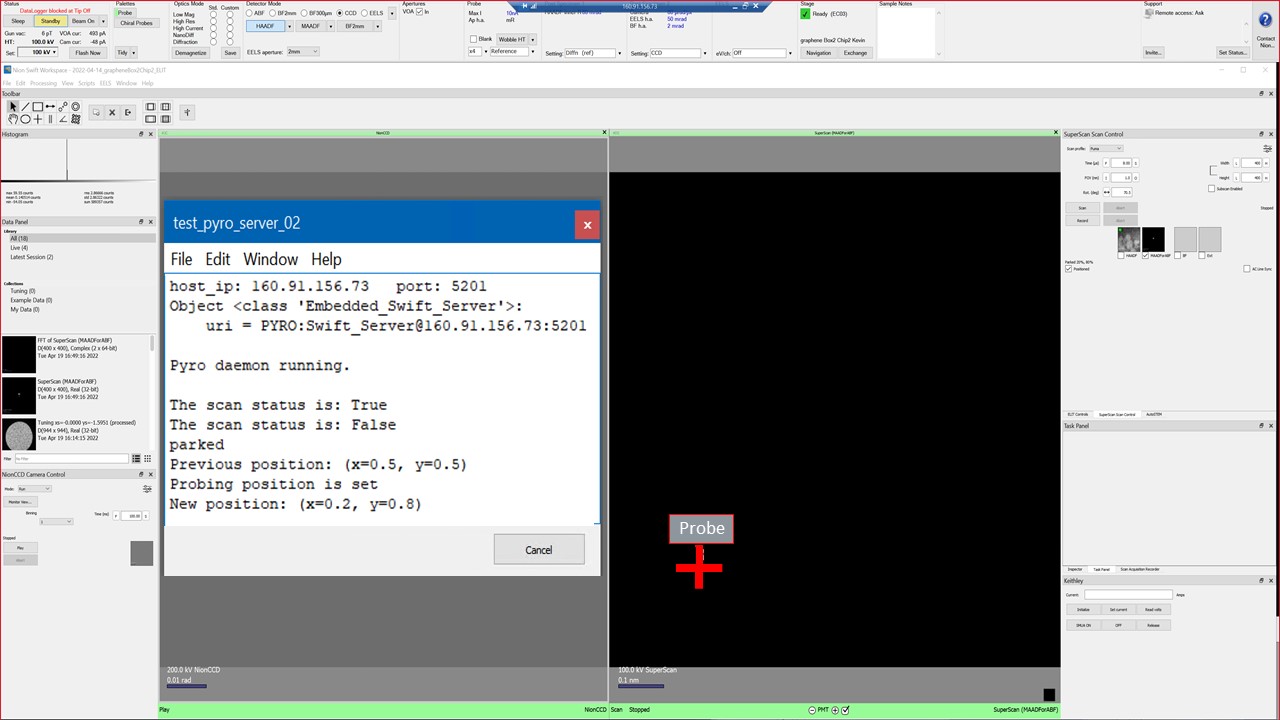}

\caption{
Swift GUI at U200 microscope control computer
\cite{al2022enabling}.}
\label{fig:U200_screenshot}
\end{figure}

\subsection{\textbf{Steering microscopy experiments over ORNL ecosystem}}
\label{subsec:Steering experiments over ORNL ecosystem}

 The microscopy control channel between the {U200} microscope and {DGX} computing system has successfully been tested to steer the microscopy experiments across ORNL physical infrastructure (Fig.~\ref{fig:U200_screenshot}). Several STEM Python-based controls are integrated into the scientific workflow to get the beam status, position the beam, and obtain instrument and experiment metadata and results.

The functions \textit{scan\_status} and \textit{probe\_position} \cite{al2022enabling} are the corresponding Pyro server objects on the {U200}  control computer. The microscope is steered by \textit{check\_scan.py} and \textit{probe\_position.py} Pyro client applications running on {DGX}.

\subsection{\textbf{Automated Data Transfers}}
\label{subsec:Automated Data Transfers}

We tested the data channel  over ORNL ecosystem by mounting the microscope measurements directory at U200 NAS  on DGX using SSHFS  file mounting. The authorized microscopists and automated codes seamlessly access NAS files at the mounted directory 
and utilize them in computations on the remote computing system.


\section{conclusions}
\label{sec: conclusion}

We presented a cyber framework consisting of network provisioning and software modules for integrating science instruments into a complex ecosystem. It utilizes separate network connections for control commands and measurement transfers across the ecosystem, and cross-platform  software modules using Pyro for communications, and  file mounting techniques for data transfers.

This framework has been implemented and tested, and it is now part of operations in supporting cross-facility Nion microscope workflows at ORNL.
It is currently under refinement and further development to support workflows in other areas, including autonomous chemistry laboratory and smart grid simulations.
Overall, this demonstration and initial results show the applicability of this cyber framework to genera

\bibliographystyle{ieeetr}
\bibliography{smartcompbib}

\end{document}